\documentclass[fleqn,usenatbib]{mnras}

\usepackage{newtxtext,newtxmath}

\usepackage[T1]{fontenc}

\DeclareRobustCommand{\VAN}[3]{#2}
\let\VANthebibliography\thebibliography
\def\thebibliography{\DeclareRobustCommand{\VAN}[3]{##3}\VANthebibliography}

\usepackage{graphicx}	
\usepackage{amsmath}	

\usepackage[version=4]{mhchem}
\usepackage{orcidlink}
\usepackage{physics}

\def\Teff{T_\mathrm{eff}}
\def\xb{\vb{x}}
\def\yb{\vb{y}}

\title[
    Semi-supervised WD Spectral Classification
]{
    Semi-supervised Spectral Classification of DESI White Dwarfs by Dimensionality Reduction
}

\author[
    X. Byrne et al.
]{
    Xander Byrne
    \orcidlink{0000-0001-9488-238X},$^{1}$\thanks{E-mail: xbyrne@ast.cam.ac.uk}
    Amy Bonsor
    \orcidlink{0000-0002-8070-1901},$^{1}$
    Laura K. Rogers
    \orcidlink{0000-0002-3553-9474},$^{1}$
    Christopher J. Manser
    \orcidlink{0000-0003-1543-5405}$^{2,3}$
\\
$^{1}$Institute of Astronomy,
University of Cambridge,
Madingley Road,
Cambridge CB3 0HA,
UK\\
$^{2}$Astrophysics Group,
Department of Physics,
Imperial College London,
Prince Consort Rd,
London,
SW7 2AZ,
UK\\
$^{3}$Department of Physics,
University of Warwick,
Coventry CV4 7AL,
UK
}

\date{Accepted XXX. Received YYY; in original form ZZZ}

\pubyear{2024}

\begin{document}
\label{firstpage}
\pagerange{\pageref{firstpage}--\pageref{lastpage}}
\maketitle

\begin{abstract}
As a new generation of large-sky spectroscopic surveys comes online, the enormous data volume poses unprecedented challenges in classifying spectra. 
Modern unsupervised techniques have the power to group spectra based on their dominant features, circumventing the complete reliance on training data suffered by supervised methods.
We outline the use of dimensionality reduction to generate a 2D map of the structure of an intermediate-resolution spectroscopic dataset.
This technique efficiently separates white dwarfs of different spectral classes in the Dark Energy Spectroscopic Instrument's Early Data Release (DESI EDR), identifying spectral features that had been missed even by visual classification.
By focusing the method on particular spectral regions, we identify white dwarfs with helium features at 90 per cent recall, and cataclysmic variables at 100 per cent recall, illustrating rapid selection of low-contamination samples from spectroscopic surveys.
We also demonstrate the use of dimensionality reduction in a supervised manner, outlining a procedure to classify any white dwarf spectrum in comparison with those in the DESI EDR.
With upcoming surveys promising tens of millions of spectra, our work highlights the potential for semi-supervised techniques as an efficient means of classification and dataset visualisation.
\end{abstract}




\section{Introduction}

A suite of upcoming large-sky spectroscopic surveys promises to enormously increase the number of sources with intermediate-resolution spectroscopy, enabling advances across all areas of astronomy.
These surveys include:
\begin{itemize}
\item
\textbf{4MOST}
(4-metre Multi-Object Spectroscopic Telescope; \citealt{4most}): nine surveys which aim to obtain accurate radial velocities and chemical abundances for millions of Galactic stars, probe the history of supermassive black hole accretion through a spectroscopic sample of about 1 million active galactic nuclei, and obtain redshifts for 10-20 million emission line galaxies;
\item
\textbf{DESI}
(Dark Energy Spectroscopic Instrument; \citealt{desii, desiii}), which plans to study the growth of structure in the Universe through observations of baryon acoustic oscillations and redshift-space distortions, and trace the dark matter distribution by obtaining redshifts for 30 million galaxies and quasars at $1.0\lesssim z \lesssim 3.5$.
When observations of these faint targets are hindered by moonlight or poor observing conditions over Iolkam Du'ag / Kitt Peak, the targeting switches to surveys of Milky Way stars and bright galaxies;
\item
\textbf{SDSS-V}
(Sloan Digital Sky Survey V; \citealt{sdssv}):
an all-sky survey from two $2.5\,\text{m}$ telescopes (one in each hemisphere), SDSS-V aims to record multi-epoch spectra for over six million targets, including five million Galactic stars;
\item
\textbf{WEAVE}
(William Herschel Telescope Enhanced Area Velocity Explorer; \citealt{weave}): a northern-hemisphere complement to 4MOST, WEAVE will record accurate spectroscopic velocities of Galactic disk and halo stars, IFU H~\textsc{i} maps of $\sim10^4$ low-redshift galaxies, and wide-area views of galaxy clusters.
\end{itemize}
These surveys will all achieve first light by early 2025; SDSS-V's first data release (SDSS DR19; \citealt{sdssdr19}) and DESI's early data release (DESI EDR; \citealt{desiedr}) were delivered in 2023.
As these spectroscopic surveys begin returning enormous quantities of data -- and with the exabyte-scale Legacy Survey of Space and Time (LSST; \citealt{lsst}) on the horizon -- automated techniques will become absolutely necessary to extract scientific results from the vast wealth of data collected.

Entries in astronomical datasets often have high dimensionality.
They may be image cutouts of hundreds of pixels, light curves with thousands of measurements, or spectra with thousands of fluxes.
Any \textit{structure} in a dataset -- such as clusters or sequences -- can be investigated quantitatively by representing each entry as a vector in a very high-dimensional space: a dataset of $N$ spectra with $D$ flux bins can be thought of as a set of $N$ points in $D$-dimensional data space.
Visualisation of datasets in more than two or three dimensions is challenging.
However, whereas a $D$-dimensional data point might ostensibly live in $\mathbb{R}^D$, it is very likely that every point in the dataset is located on or near a much lower-dimensional submanifold within this space.
In a spectrum, the flux in a given bin will usually be similar to the flux in adjacent bins.
Similarly, two different sources of the same astrophysical type (e.g.\ two stars of the same spectral classification) will have similar spectra, and will thus be nearby in the data space.
This massively reduces the effective dimensionality of the dataset, compared to the dimensionality of the raw data points themselves.
This principle has motivated the development of several \textit{dimensionality reduction} techniques, including Principal Component Analysis (PCA; \citealt{pca}), self-organising maps \citep{kohonen90}, diffusion maps \citep{coifman06, lafon06}, Locally Linear Embedding (LLE; \citealt{roweis00}), $t$-distributed Stochastic Neighbour Embedding ($t$SNE; \citealt{tsne}), and Uniform Manifold Approximation and Projection (UMAP; \citealt{umap}).
These methods calculate a two-dimensional map of the dataset -- known as an \textit{embedding} -- in which the distances between pairs of nearby points in the data space are preserved as far as possible.

Dimensionality reduction techniques have found widespread use in a variety of areas of astronomy, including the analysis of low-resolution spectra.
\citet{boroson92} use PCA to explore correlations between the physical parameters of 87 low-redshift quasars observed in a small spectroscopic campaign.
\citet{richards09} employ PCA alongside diffusion maps to predict redshifts for several thousand low-redshift galaxy spectra observed by SDSS DR6.
\citet{hawkins21} use $t$SNE to reduce the dimensionality of a set of low-resolution ($R\equiv \lambda / \Delta \lambda \sim750$) optical spectra of stars in the Hobby-Eberly Telescope Dark Energy Experiment (HETDEX; \citealt{gebhardt21}) to estimate a star's effective temperature ($\Teff$), gravitational field strength ($\log g$), and metallicity, finding 416 metal-poor candidate stars.
\citet{kao24} apply UMAP to low-resolution ($R\sim50$) Gaia DR3 XP coefficients of white dwarf (WD) spectra, isolating a group of 465 objects of which 90 are known \textit{polluted} WDs, showing metal absorption features; the remaining 375 candidates are subject to an ongoing high-resolution spectroscopic campaign.
Dimensionality reduction has however not been extensively applied to datasets of higher-resolution spectra, perhaps due to the much higher dimensionality of spectra with many more wavelength bins.

WDs are the final state of the $\gtrsim97$ per cent of stars with a zero-age mass of less than $9$--$12\,\mathrm{M}_\odot$ \citep{lauffer18, althaus10, althaus21}.
Without internal nuclear fusion, these stellar cinders gradually cool on well-characterised timescales, making them important tracers of the evolution and assembly of the Galaxy (e.g., \citealt{winget87, tremblay14}).
The spectrum of a WD usually deviates from that of a black body due to absorbing species in its atmosphere.
With a typical mass of $\sim0.6\,\mathrm{M}_\odot$ and radius of $\sim \mathrm{R}_{\earth}$, the extreme gravitational field gives WDs a strongly stratified structure.
Metals diffuse down through WD atmospheres on timescales much shorter than their cooling age \citep{schatzman45, paquette86, koester09, wyatt14}, leaving thin photospheres dominated by H and/or He.
Most WD spectra show absorption features due to these light elements, and a WD is classified respectively as DA or DB if spectral features of neutral H or He are present.
Hotter WDs may show ionised  He~\textsc{ii} features (DO).
Cooler WDs cease to excite atomic transitions below particular temperatures, ($\lesssim5000\,\text{K}$ for H; $\lesssim 11000\,\text{K}$ for He) and the absorption lines gradually fade away, giving a featureless spectrum (DC).
In certain WDs, the convection zone reaches deep into the interior and dredges up carbon from the inert core, imprinting molecular \ce{C2} Swan bands in the spectrum (DQ; \citealt{fontaine84, koester82, koester20, blouin23}).
In other cases, instabilities in remnant planetary systems can cause planetesimals to be tidally disintegrated and accreted by WDs, lending metal absorption features to the spectrum \citep{bonsor11, frewen14, mustill18, maldonado20}.
These so-called polluted WDs provide unique insights into the bulk composition and geology of rocky exoplanets.
In the absence of other features, such metal-line WDs are classified DZ.
If, say, both H features and metal features are both present, the WD is instead classified as DAZ or DZA, depending on which features are strongest.
Similarly, WDs may have more auxiliary classifications, which are listed in order of prominence in the spectrum.
Thus a DBZA shows strong neutral He features, weaker metal features, and even weaker H features. For convenience, we will refer to WDs with a particular \textit{primary} classification as, for example `DA*', to include DA, DAB, DAZ, etc.
Additionally, we use the broader classification `D*A' to include WDs with primary \textit{or} secondary classification, inclusive of DA, DAB, DZA, etc.
D*A therefore encompasses any WD whose spectrum contains visible Balmer features.
Similarly, D*Z describes any polluted WD, including those with other stronger spectral features.

Following the release of the DESI EDR \citep{desiedr}, the spectra of 3673 WD candidates were visually inspected and classified by \citet{manser24}, spectroscopically confirming 2706 WDs, of which 1400 had not been previously observed spectroscopically.
The higher resolution of DESI ($R\sim2500$--$5000$) enables the identification of spectral features not visible in lower-resolution spectra.
Of the 152 polluted WDs identified, 121 were newly discovered or previously not classified as polluted.
However, visual classification requires significant amounts of expert time, and substantial increases in target numbers in the future ($\sim70\,000$ WD candidates from the full DESI survey; \citealt{cooper23}) all but necessitates the use of automated methods.

One class of methods for automatically classifying and analysing WD spectra that has achieved success is supervised machine learning (e.g., \citealt{yang20, tan23, garciazamora23, vincent23, vincent24}).
However, a drawback of supervised machine learning techniques is the reliance on the training set used to train the model.
The `imbalanced learning problem' describes the well-documented difficulties of training supervised classifiers on datasets with large class imbalances (e.g., \citealt{he09, johnson19}).
Such class imbalances are inevitable in WD datasets as some spectral types are naturally rarer than others: the DESI EDR contained 1958 DAs, but only one DO \citep{manser24}.
Although this may be mitigated through the addition of synthetic spectra to the training set, any biases regarding the generation of the spectra or the modelling of instrumental effects would simply be learned by the trained model.
In addition to class imbalance, incorrect labels in the training set can severely limit the accuracy of the resulting model (e.g., \citealt{frenay14}).
Unsupervised procedures such as dimensionality reduction do not rely on an external training set; they merely reveal structures present within the dataset itself.

In this work, we apply dimensionality reduction -- specifically, $t$SNE -- to intermediate-resolution WD spectra from the DESI EDR, outlining the method's ability to classify WD spectra in an semi-supervised way.
Dimensionality reduction itself is unsupervised, but the analysis of the resulting embedding uses the visual classifications of \citet{manser24}.
We emphasise that these methods could equally be applied to sets of main-sequence stars, quasars, galaxies, or any other set of sources targeted by the aforementioned spectroscopic surveys.
The paper proceeds as follows.
In Section~\ref{sec:methods}, we describe the dataset and dimensionality reduction in more detail, as well as important preprocessing considerations.
In Section~\ref{sec:results}, we demonstrate this method's ability to identify clusters and sequences in the DESI EDR WD catalogue.
We also outline a way to incorporate prior knowledge about the locations of distinctive spectral lines, and a method of using dimensionality reduction in a supervised manner to classify spectra external to the catalogue.
Section~\ref{sec:discussion} discusses the application of this method as an aid in the classification of large spectroscopic WD datasets.
Section~\ref{sec:conclusions} summarises our work.

\section{Methods}
\label{sec:methods}

\subsection{Data}

The DESI EDR contains $N=3673$ WD candidates from the catalogue compiled by \citet{gentilefusillo19} for which the exposures include a median signal-to-noise ratio $>0.5$ in at least one of DESI's three spectral arms \citep{manser24}.
These publicly available\footnote{
    \url{https://data.desi.lbl.gov/public/index.html}
} exposures were obtained and stacked by standard inverse variance weighting.
The spectra span $3600$--$9824\,\text{\AA}$, with a wavelength spacing of $\Delta \lambda = 0.8\,\text{\AA}$, giving $D=7781$ flux bins for each spectrum.
All spectra within DESI have an identical wavelength axis.

\subsection{Dimensionality reduction}

A spectrum can be represented by a high-dimensional vector, with each component of the vector corresponding to the flux in a given wavelength bin.
Spectra produced by the DESI pipeline have 7781 bins, so they can be mapped one-to-one to 7781-dimensional vectors.
These vectors are not uniformly-distributed in $\mathbb{R}^{7781}$;
spectra with shared spectral features are clustered together, as their vector components corresponding to wavelength bins near these features will take similar values.
For example, the 3703rd component corresponds to a wavelength of $6562.4\,\text{\AA}$, very close to the H$\,\alpha$ line.
The (normalised) spectra of DA WDs will have lower values in their 3703rd component than spectra without any H absorption features.
DAs will hence -- in this component, and by extension other components -- be close together.
Additionally, in the absence of artefacts or large noise, adjacent components of a given spectrum are likely to have similar values.
The components of spectra are thus highly intercorrelated.
Dimensionality reduction techniques exploit these redundancies in datasets, seeking a low- (usually two-) dimensional representation of a dataset's structure.

A popular dimensionality reduction technique is $t$-distributed Stochastic Neighbour Embedding ($t$SNE; \citealt{tsne}).
Briefly, $t$SNE attempts to map a set of high-dimensional vectors $\xb_1, \xb_2, \dots, \xb_N \in \mathbb{R}^D$ into a set of two-dimensional embedding vectors $\yb_1, \yb_2, \dots, \yb_N \in \mathbb{R}^2$, in such a way as to preserve the `similarity' between each pair of data points.
The details of how this is achieved are given in Appendix~\ref{sec:app} (see also \citealt{tsne}).
Many other dimensionality reduction techniques, such as Uniform Manifold Approximation and Projection (UMAP; \citealt{umap}) and LargeVis \citep{largevis} can be shown to belong to the same family of procedures, but with differing definitions of similarity (see Appendix C of \citealt{umap}).

\subsection{Data preprocessing}
\label{sec:preprocessing}

\begin{figure*}
\centering
\includegraphics[width=0.97\textwidth]{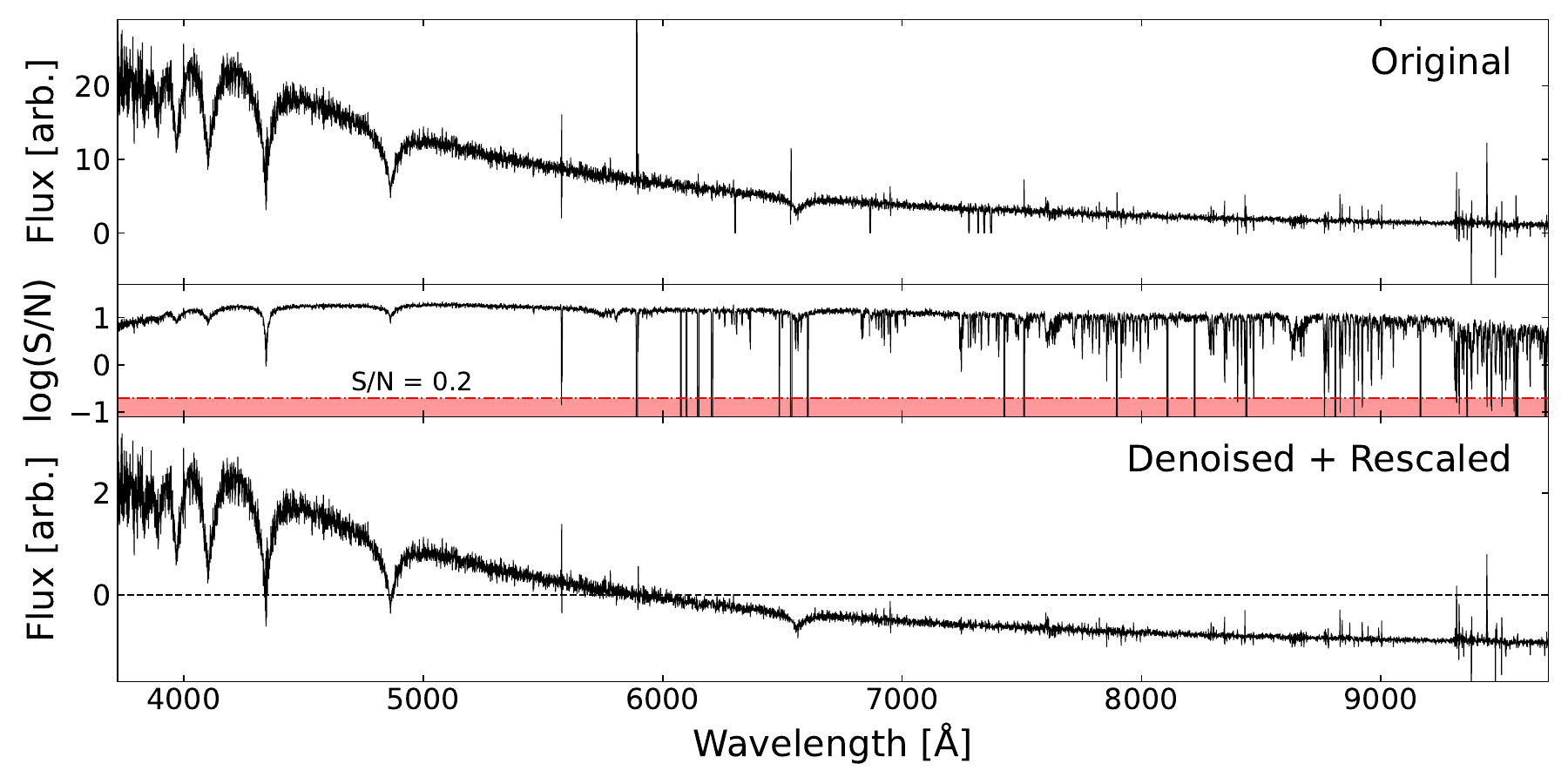}
\caption{
    Preprocessing stages, as illustrated on the cherry-picked DESI EDR spectrum of WD J170114.72+760207.16.
    The upper panel shows the raw spectrum, with several artefacts.
    The second panel shows that the signal-to-noise ratio is very low near many of these artefacts; where it falls below 0.2, the pixels are interpolated.
    The spectra are then rescaled to zero mean and unit variance, as shown in the lower panel.
    Major artefacts have been removed, though certain imperfect sky subtraction and telluric features remain, particularly above $7500\,\text{\AA}$.
}
\label{fig:preprocessing}
\end{figure*}

Before applying dimensionality reduction, it is of paramount importance to preprocess the spectra.
The spectra show a range of absolute scales: the median fluxes span three orders of magnitude.
A na\"{i}ve application of dimensionality reduction to the raw data would cause two identical WDs at different distances (and hence different brightnesses) to be embedded far away from each other.
As we are interested in the intrinsic properties of the WDs in this sample, rather than their distance from Earth, the spectra must be normalised before dimensionality reduction is applied.
Additionally, some spectra show large spikes, due to cosmic rays, detector artefacts, or imperfect sky subtraction (see Fig.~\ref{fig:preprocessing}).
Interpreted as vectors, this corresponds to some components of the vector being very large, and as a result the vector would be far away in data space from where it would be without the artefacts.

The preprocessing we apply to the data is illustrated in Fig.~\ref{fig:preprocessing}.
Artefacts usually correspond to pixels with very low signal-to-noise: pixels with $\mathrm{S}/\mathrm{N}<0.2$ are linearly interpolated.
Fig.~\ref{fig:preprocessing} shows that major artefacts are removed by this step, though some sky lines and telluric features remain.
Overzealous removal of artefacts could remove genuine spectral features, particularly weaker ones.
We found that the interpolation described improves the subsequent dimensionality reduction as compared to, say, median boxcar smoothing.
Following the removal of artefacts, the spectra are normalised to zero mean and unit variance, to account for the fact that WDs at different distances will show different absolute fluxes.
This particular normalisation has been used in previous work in the data-driven spectroscopic analysis of WDs (e.g., \citealt{vincent23}), and we found this gave better results than many other normalisation strategies.

In its simplest form, dimensionality reduction treats all components of the vectors equally.
However, some wavelengths are of more significance than others, namely those of absorption and emission features.
To `focus' the technique on particular spectral lines, we crop the spectrum to an appropriate window before applying dimensionality reduction.
This additional preprocessing step is explored in Section~\ref{sec:zoom}.

\section{Results}
\label{sec:results}

\subsection{Applying \textit{t}SNE to the full spectra}
\label{sec:full_spectra}

\begin{figure*}
    \centering
\includegraphics[width=0.5\textwidth]{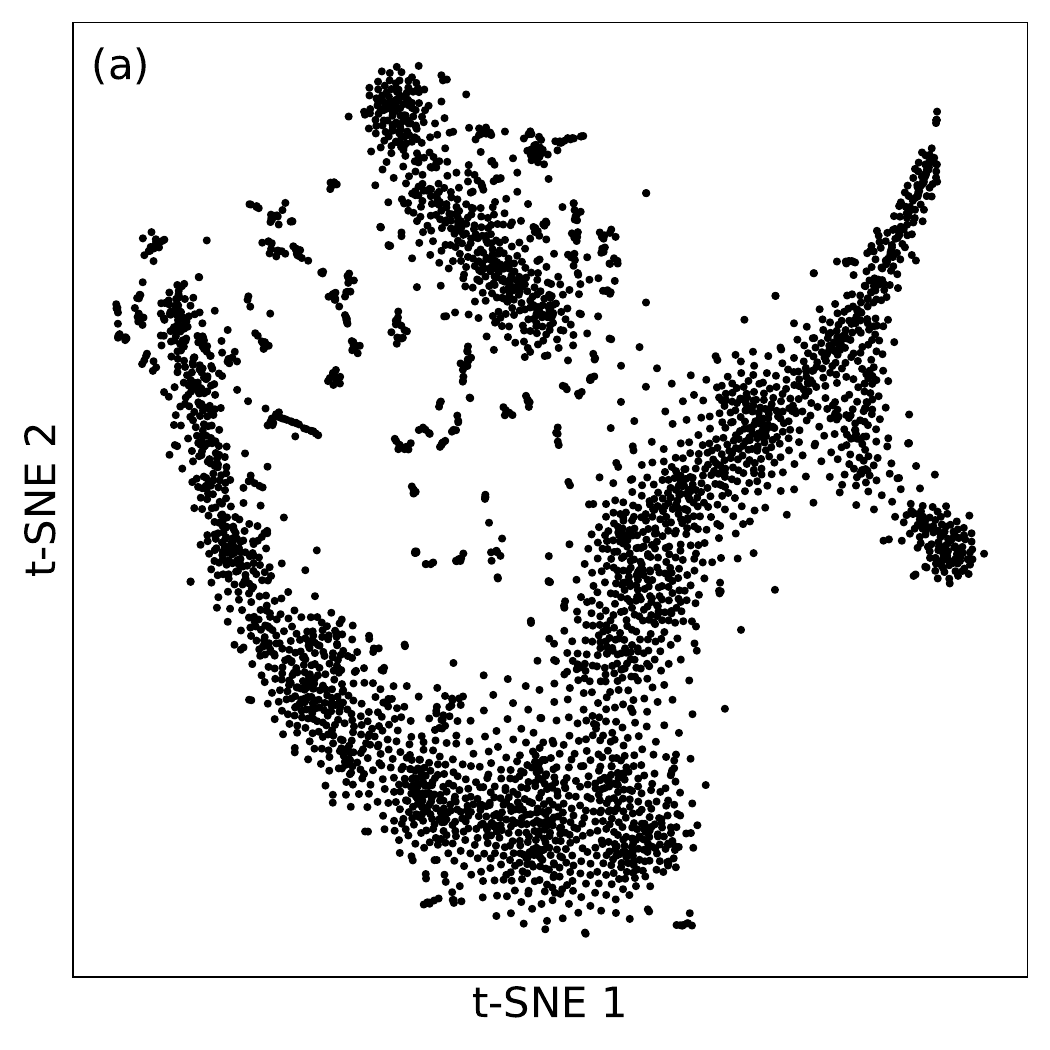}
\includegraphics[width=0.98\textwidth]{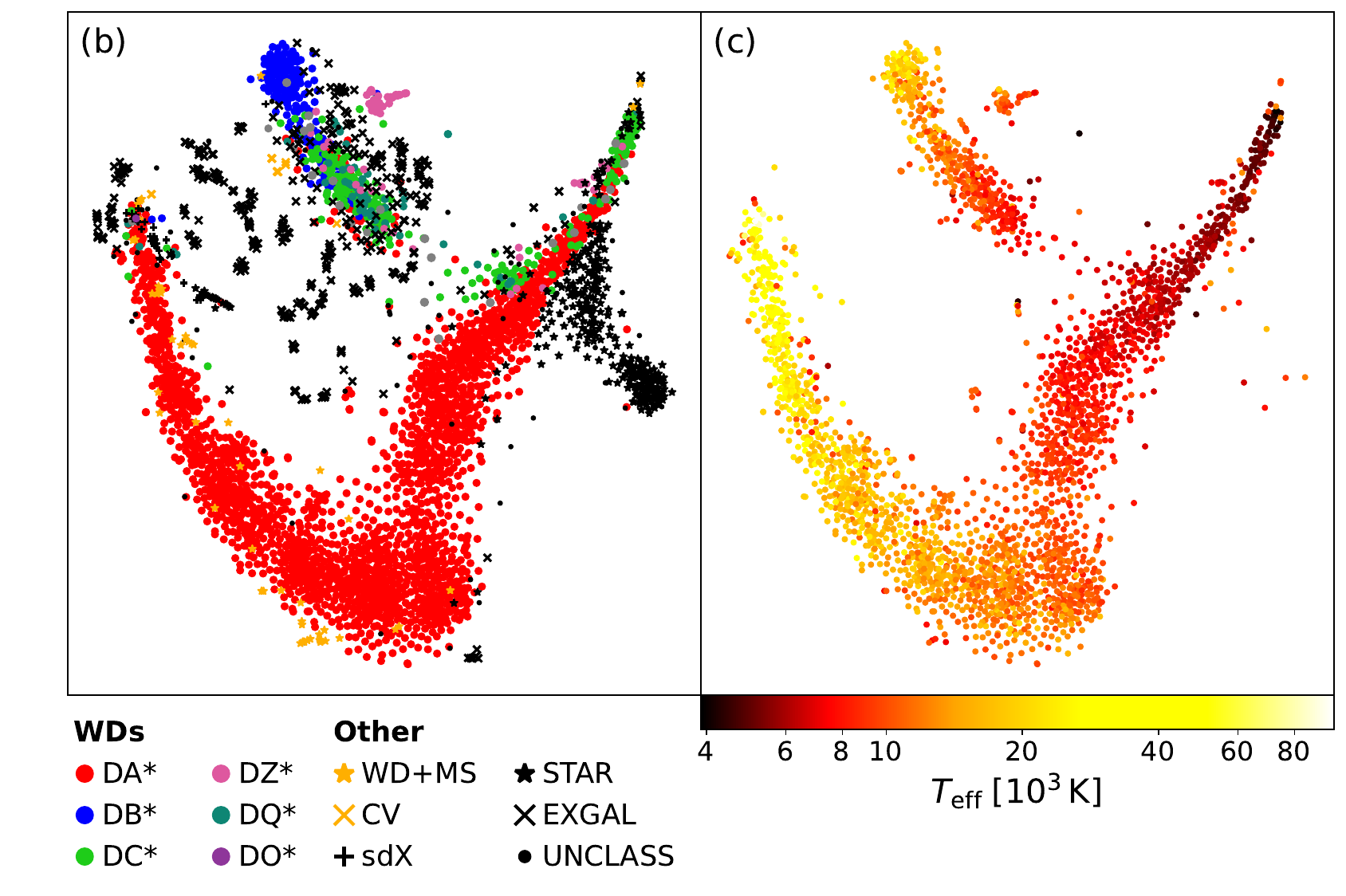}
\caption{
    (a) $t$SNE embedding of DESI EDR WD candidate spectra.
    The embedding aims to reflect the pairwise distances between the high-dimensional spectra in a two-dimensional space; as such the axes are arbitrary.
    (b) The embedding is colour-coded according to the visual spectral classification of \citet{manser24}.
    For WDs, the colour corresponds to primary spectral type: DA*s (inc.\ DAZs, etc.) in red; DB*s in blue, and so on.
    Other sources not corresponding to individual WDs have different symbols (see key).
    The main feature of the embedding is the sequence of DA*s (red), though several other clusters are clear (see text).
    (c) The embedding is colour-coded by effective temperature, according to the hydrogen-atmosphere WD model which best fits the sources' Gaia photometry \citep{gentilefusillo19}.
    The DA sequence extends from hotter WDs on the left (among which lies the singular DO), around to cooler WDs at the top right.
    Note that objects with a low probability of being a WD ($P_\mathrm{WD} < 0.75$; see \citealt{gentilefusillo15}) are not assigned a temperature \citep{gentilefusillo19}.
}
\label{fig:full_spectra}
\end{figure*}

Dimensionality reduction was applied to the spectra of the WDs in the DESI EDR WD catalogue, using $t$SNE.
The dimensionality reduction takes an average of $5.5\,\text{s}$ to process the entire dataset.
The resulting embedding is shown in Fig.~\ref{fig:full_spectra}(a).
Each point in this embedding corresponds to an individual spectrum, and several clusters and sequences are apparent.
There is a long V-shaped sequence stretching from the left, to the bottom, to the upper right of the embedding; it turns off into a cluster towards the centre right.
A shorter secondary sequence extends from the top of the embedding towards the centre.
Several strings and clumps are found in between and around these two sequences.

The nature of these structures is elucidated by Fig.~\ref{fig:full_spectra}(b) and (c), showing respectively the primary spectral class (according to \citealt{manser24}) and effective temperature (according to H-atmosphere model fitting of \citealt{gentilefusillo19}).
Several interesting features include:
\begin{itemize}
\item
The V-shaped sequence consists of largely DA* WDs;
\item
The secondary sequence in the upper part of the frame contains DB*s, DZ*s, DQs, and DCs;
\item
The WDs in the DA sequence have been sorted approximately according to effective temperature, with hotter WDs towards the left of the sequence;
\item
DZ*s are found in various places\footnote{
    \citet{kao24} also find that DZs with different characteristics (primarily $\Teff$) are scattered by dimensionality reduction into multiple regions.
}: (i) an island near to the secondary sequence, (ii) towards the cool end of the DA sequence, (iii) amongst the secondary sequence;
\item
Extragalactic sources (`$\times$' markers) are mostly located in strings and small clumps around the secondary sequence;
\item 
Subdwarfs (`$+$' markers) are largely found towards the hot end of the DA sequence;
\item
The majority of main-sequence stars (black star icons) are found towards the centre right, though their cluster merges somewhat with the cool end of the DA sequence (cf.\ fig.~1 of \citealt{eisenstein06}).
A few are also scattered elsewhere.
\end{itemize}

Although the spectral classes are known here, owing to the visual inspection campaign of \citet{manser24}, dimensionality reduction would nonetheless provide a valuable analysis if the classes were \textit{a priori} unknown, as will be the case with upcoming spectroscopic surveys.
For example, it is well-known that approximately 80 per cent of WDs are DAs.
As such, when applying dimensionality reduction to a WD dataset, one would implicitly identify the largest cluster as corresponding to mostly DAs.

That dimensionality reduction is capable of separating different classes of WD into groups in this way is ultimately due to similarities between spectra of the same spectral class, and differences between different classes.
Expressed as vectors, DA spectra (say) are much closer to vectors of other DA spectra of similar effective temperature, than to DB spectra.
Note that this would not be true if the spectra had not been preprocessed (see Section~\ref{sec:preprocessing}): closer WDs would have brighter spectra than more distant WDs of the same spectral type, and hence spectral vectors with larger magnitudes.
The spectral vectors would therefore be very far away from each other, and hence also embedded far apart, despite being similar astrophysical objects.
The same would be true of spectra with very large artefacts.
Preprocessing is thus crucial in removing any irrelevant aspects of the data.

The H-atmosphere model fitting of \citet{gentilefusillo19} suggests that the V-shaped sequence transitions smoothly from $\Teff \gtrsim 80000\,\text{K}$ through to $\Teff \lesssim 4000\,\text{K}$.
We see that the variation in the tilt of the spectral black-body continuum -- as well as perhaps second-order features such as the strength of the Balmer lines -- has been converted by dimensionality reduction to variation along this sequence.
The hottest WD in this sequence has a very similar spectrum to the second-hottest, so the two are embedded close together; the second-hottest has a similar spectrum to the third-hottest; etc.
Anomalies of intermediate temperature at opposite ends of the sequence (see Fig.~\ref{fig:full_spectra}(c)) are largely main-sequence stars and other objects that have been erroneously assigned an ill-fitting WD temperature.
The fact that $\Teff$ ``spans'' the embedding identifies this parameter as that which describes a spectrum to first order: the parameter which primarily determines the spectrum is the temperature.
The effects of $\Teff$ in the dimensionality reduction are discussed in more detail in Section~\ref{sec:discteff}.

The transitions undergone by WDs as they gradually cool are also borne out in the embedding (Fig.~\ref{fig:full_spectra}(b)), though some subtleties emerge as a consequence of the classification system itself.
As the temperature of a DA falls below around $\Teff\approx 5000\,\text{K}$, hydrogen transitions are no longer excited, the Balmer lines fade away, and the DA becomes a DC.
However, the transition is continuous: Balmer lines do not suddenly disappear below some sharp temperature cutoff.
As such, the distinction between DA and DC is a fuzzy one, and whether a cool hydrogen-atmosphere WD spectrum is classified as DA or DC ultimately depends on whether a human classifier can subjectively identify H features against noise.
This task is not trivial, but we argue that it is ultimately arbitrary: there is negligible physical difference between a DA with very weak Balmer lines, and a hydrogen-atmosphere DC.
Unable to distinguish between these two arbitrarily similar classes, dimensionality reduction thus blends the low-temperature DAs with the DCs, in a sense including the \textit{strength} of spectral features in the classification, rather than the binary presence or absence of features.

A DC WD, as labelled by visual classification campaigns, is not necessarily one with a pure black-body spectrum; it is merely one where any spectral features are at a level smaller than the noise.
It is therefore likely that higher-signal-to-noise spectra would reclassify some of the putative DCs in the DESI EDR WD catalogues.
Conversely, it is possible that noise features at the location of a spectral feature in the spectrum of a genuine DC might fool a human classifier into labelling it otherwise.
Rather than exhibiting a weakness of the technique, we argue that a cool DA is objectively more similar to a H-atmosphere DC, than it is to a much hotter DA with strong Balmer features.
It is therefore more natural that cool DAs should be projected near to DCs.
Dimensionality reduction faithfully reflects this subtlety by projecting the cool DAs with weak absorption features nearby to DCs with no visible features.
Similarly, the hot end of the DA sequence is somewhat blended with other classes of very hot objects, most commonly subdwarfs.
Nonetheless, the central stretch of the DA sequence shows that dimensionality reduction very precisely identifies WD Balmer lines, with between 92 and 98 per cent of this midsection classed as D*As or WD-main-sequence binaries by \citet{manser24} (depending on how liberally one defines the midsection).
This technique therefore identifies WDs with clear Balmer features at high precision, and edge cases are faithfully presented as such.
While not necessarily assigning a strict classification to these transitional WDs, such a classification is physically not very significant anyway. 

Similar trends are borne out in the secondary sequence, which shows DB*s at the top of the embedding transitioning into a mixture of DCs and DQs.
The transition from DB to DC has a very similar physical origin to the DA-to-DC transition discussed in the previous paragraph, but at higher temperature, as seen in Fig.~\ref{fig:full_spectra}(c).
The DC transition temperatures are in line with theoretical predictions of approximately $5000\,\text{K}$ and $11000\,\text{K}$ for H- and He-dominated atmospheres respectively.
Aside from forming a DC, a cooling He-atmosphere WD may begin dredging carbon from the core, giving rise to a DQ.
As this cooling occurs, He transitions fail to be excited and the He lines of a DB fade away.
As with Balmer lines, He lines and Swan bands can be arbitrarily weak, blurring the boundaries between cool DBs, DQs, and DCs.
That these three spectral types are all projected together in the secondary sequence at the top of the embedding reflects the smooth transition between them, as with the H-atmosphere WDs discussed above.
Indeed, Fig.~\ref{fig:full_spectra}(c) shows that the DCs in this part of the embedding have temperatures\footnote{
    Although the temperatures shown in Fig.~\ref{fig:full_spectra}(c) are based on H-atmosphere models, the He-atmosphere model temperatures are very similar.
} well in excess of the DA-to-DC transition temperature of $\approx 5000\,\text{K}$.
These DCs therefore must have He-dominated atmospheres: if they were H-dominated then Balmer features would be visible at these temperatures.
These observations therefore suggest that the two main features of the embedding correspond respectively to H- and He-dominated atmospheres.

Briefly, we discuss the presence of a small number of DA*s at the cool end of the He-atmosphere sequence (see Fig.~\ref{fig:full_spectra}(b), just above centre), of which there are 43.
Of these, 24 are classified by \citet{manser24} as DAH, DAe, DAP, or some combination of these secondary classifications.
These may have been found in the He-atmosphere sequence as the broad Swan bands of a DQ are mimicked by the Zeeman-split Balmer features of a DAH, or broad emission features of a DAe, or other unidentified features in DAPs.
The remaining 19 are either pure DAs, DABs, DAZs, or some combination of these classes.
These objects are invariably warm ($\Teff \gtrsim 7000\,\text{K}$) and have weak Balmer lines, making them appear similar to DCs of similar temperatures.
It is thus understandable that these spectra of primary classification DA are projected near to these warm (thus He-atmosphere) DCs.
They may be He-dominated WDs between the threshold temperatures for H and He absorption excitation ($5000$--$11000\,\text{K}$), such that Balmer features dominate over He lines even though the H abundance is lower than He.

It is somewhat surprising that there is a small gap between the two sequences.
Based on the observations outlined above, this region should be populated with DCs of $\approx7000\,\text{K}$, but in fact there are very similar DCs of around this temperature on either side.
It is possible that the gap between this region and the secondary sequence containing the rest of the He-atmosphere WDs is simply due to poor sampling around this temperature.
Indeed, \citet{manser24} also find a gap in the distribution of $\mathit{BP}-\mathit{RP}$ colours of DCs in the DESI EDR (see Section~\ref{sec:colourcolour}).
With the larger number of objects that will be provided by the full releases of upcoming spectroscopic surveys, this gap may be filled in.

\begin{figure}
\includegraphics[width=0.47\textwidth]{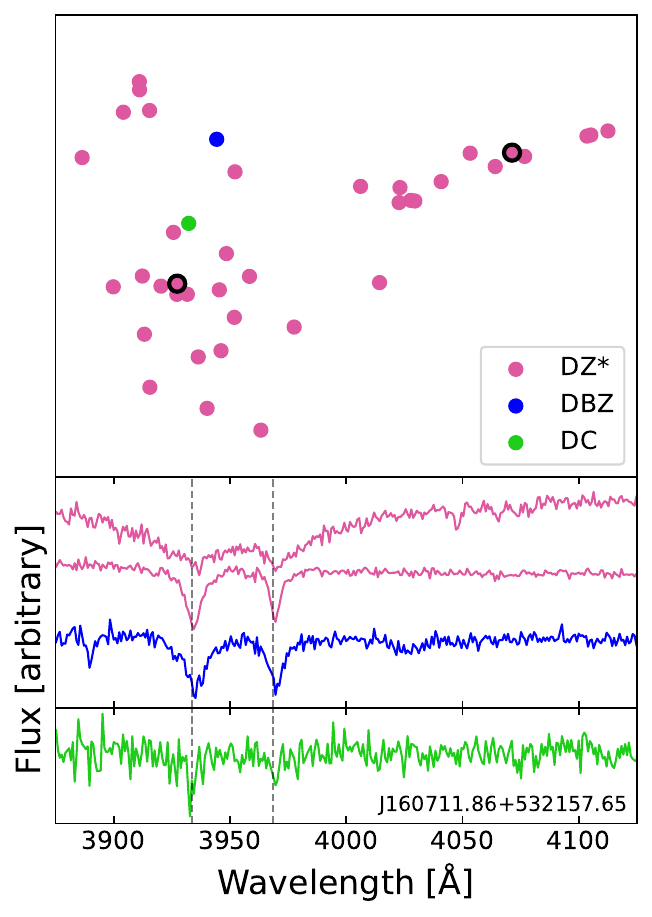}
\caption{
    Zoomed-in view of the upper centre of the embedding shown in Fig.~\ref{fig:full_spectra}; spectra of four WDs.
    Except for two objects classified by \citet{manser24} as DBZ and DC, all the objects on this island have primary classification DZ.
    Two of these DZs (highlighted) have their spectra shown in the second panel, along with the DBZ spectrum.
    All three spectra show strong absorptions due to Ca~\textsc{ii} (dashed vertical lines).
    The final panel shows the putative DC, which also shows absorption features here.
}
\label{fig:DZisland}
\end{figure}

The automated nature of dimensionality reduction can protect against human error in the visual classification of WD spectra.
Fig.~\ref{fig:DZisland} shows a zoom-in to the top of the embedding around an island populated mostly by DZ*s.
According to the visual classifications of \citet{manser24}, this island also contains a DBZ (J133305.34+325400.11) and a DC (J160711.86+532157.65).
These objects appear to have been grouped together owing to the presence of Ca~H and K lines in all the spectra, including the putative DC.
The presence of Ca absorption lines in the DC's spectrum suggests that this is in fact a DZ; indeed \citet{kleinman13} classify this source as a DZ based on its SDSS DR7 spectrum.
The bottom panel of Figure~\ref{fig:DZisland} shows the absorption lines to be quite weak, and would probably not have been noticeable without zooming in.
While large visual classification campaigns are reliable, this example illustrates that errors are nonetheless possible, and that such errors can be easily identified with automated methods such as dimensionality reduction.
This particular case also illustrates the ability of dimensionality reduction to quickly create a high-recall sample of polluted WDs (in this case 100 per cent recall) from a spectroscopic dataset.

Finally, we note that there is no clear trend in the radial velocity of the WDs in the sample.
There is a very narrow range of redshifts in the sample: around 97\% of the DAs have a redshift within 0.005 of the mean.
The small radial velocities involved, along with the breadth of the Balmer features, lead to only small differences between spectral vectors.
Interestingly, trends in redshift \textit{are} apparent among quasars in the sample.
Along each `string' of extragalactic sources shown in the embedding, the wavelength of the Ly~$\alpha$ line gradually shifts.

\subsection{Classification using specific spectral regions}
\label{sec:zoom}

As mentioned in Section~\ref{sec:preprocessing}, dimensionality reduction treats all components of a vector -- i.e.\ all wavelengths -- equally.
However, wavelength ranges that include spectral features are of particular interest.
In this subsection we explore the use of cropping the spectrum, to a window around some particular spectral line, to see if dimensionality reduction can better separate individual spectral classes from the rest.

\subsubsection{Helium features}
\label{sec:DBs}

\begin{figure}
\centering
\includegraphics[width=0.48\textwidth]{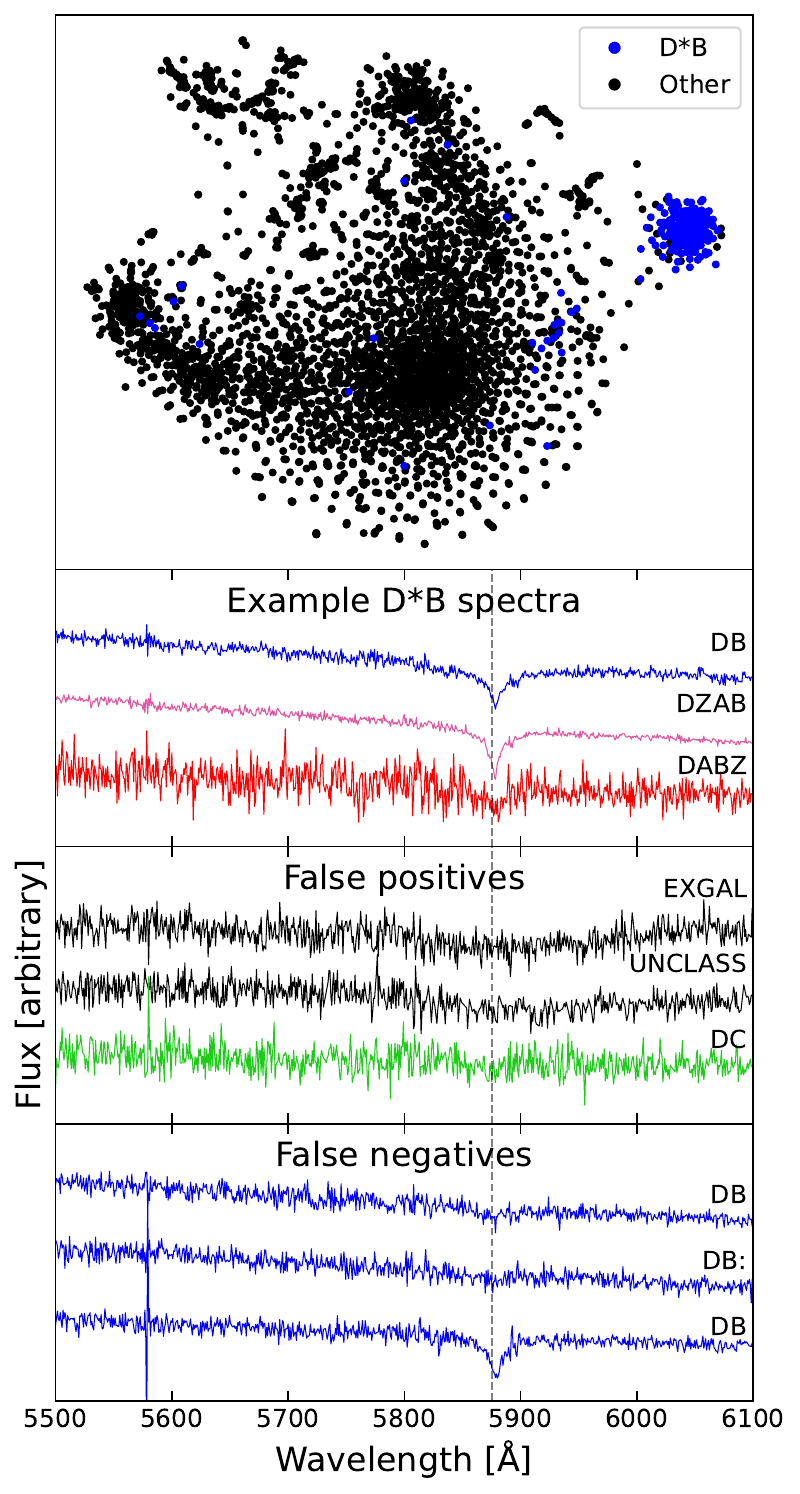}
\caption{
    Top panel: dimensionality-reduced embedding of DESI EDR spectra cropped between $5500$--$6100\,\text{\AA}$.
    An island is isolated containing about 180 objects classified as D*Bs by \citet{manser24}; some D*Bs are not on the island.
    Second panel: three spectra on the island are shown, each featuring a He absorption at $5876\,\text{\AA}$ (dashed line).
    Third panel: three objects on the island which are however not visually  classified as D*Bs.
    Shallow absorptions are just about visible.
    Bottom panel: three D*Bs \textit{not} located on the island.
    The first two spectra show very weak He features; the third spectrum shows a large artefact.
}
\label{fig:He_lines}
\end{figure}

We first crop the spectra to the range $5500$--$6100\,\text{\AA}$, a window which includes a He line at $5876\,\text{\AA}$.
The dimensionality of the spectra cropped to this wavelength range is $D=750$, much lower than $D=7781$ for the full spectra.
The result of applying dimensionality reduction to this cropped dataset is shown in Fig.~\ref{fig:He_lines}.

The technique separates an island of around 200 objects, containing the vast majority ($\approx 180/200=90$ per cent) of WDs with He lines (D*B).
The second panel of Fig.~\ref{fig:He_lines} demonstrates that these WDs are isolated primarily due to their shared $5876\,\text{\AA}$ absorption.
This shared feature causes the spectral vectors to be close together in data space when the dimensionality is confined to this subset of dimensions; this proximity is preserved by dimensionality reduction in two dimensions.

The third panel of Fig.~\ref{fig:He_lines} shows three examples of objects that are projected onto this island despite not being classed as D*Bs by \citet{manser24}.
For some of these spectra there is perhaps a very shallow or broad absorption feature, but it is difficult to tell by eye.
These spectra also appear to be relatively noisy, so these spectra may have been projected nearby due to coincidentally-similar noise.
The distinction between a genuine weak spectral feature and noise would be difficult for dimensionality reduction to ascertain, as it does not account for variations in flux between different spectral bins of the same spectrum, but rather differences in flux in the \textit{same} bin between \textit{different} spectra.

The bottom panel of Fig.~\ref{fig:He_lines} shows examples of false negatives: D*Bs which are not on the island.
The first two have very weak He absorption lines, and have been identified as DBs tentatively or through spectral features in other parts of the spectrum.
The third object has a large noise feature at $5577\,\text{\AA}$, likely a [O~\textsc{i}] line escaping imperfect sky subtraction\footnote{
    This line is also seen in the false positive DC.
}.
This noise feature also escaped removal by the preprocessing step, and has led its spectral vector to be far enough away from the other D*Bs that is has not been grouped with them.

\subsubsection{Balmer lines and cataclysmic variables}
\label{sec:CVs}

Cataclysmic variables (CVs) are binary systems in which a donor star overfills its Roche lobe and transfers matter onto a WD.
The characteristic spectral features of a CV are strong emission lines from the transferred material, which depending on the orientation of the system and the WD's magnetic field strength may be double-peaked \citep{smak69, huang72}.

\begin{figure}
\centering
\includegraphics[width=0.48\textwidth]{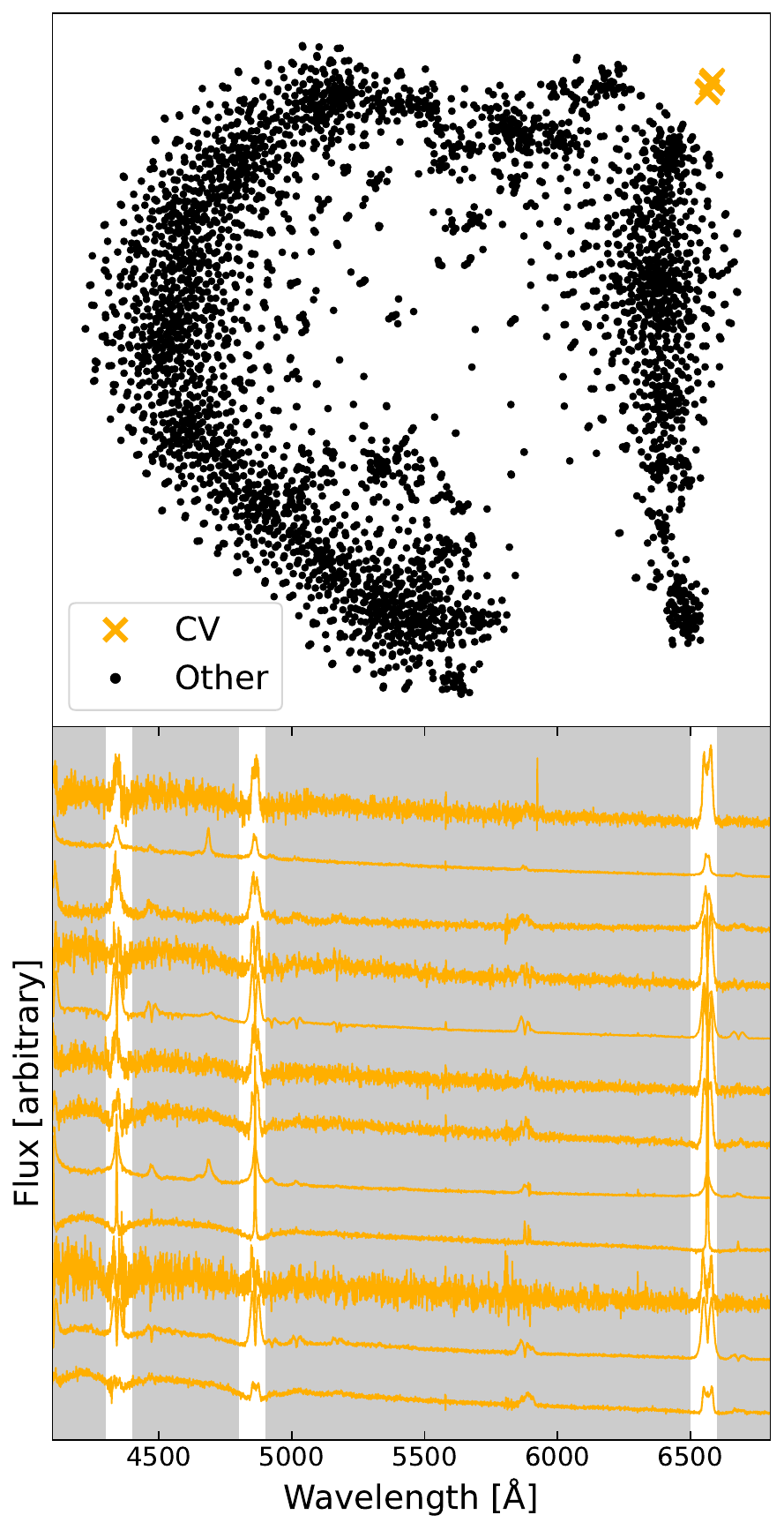}
\caption{
    Top panel: dimensionality-reduced embedding of DESI spectra cropped to three regions around H\,$\alpha$, H\,$\beta$, and H\,$\gamma$.
    An isolated island to the top right shows all 12 CVs in the sample, identified as similar to each other by shared emission features in these regions.
    Lower panel: spectra of the CVs.
    The ranges over which the spectra were cropped before dimensionality reduction is highlighted, and surrounds the strong, often double-peaked emission features.
}
\label{fig:CVs}
\end{figure}

As these unique features are most commonly seen in the Balmer series, we isolate CVs from the DESI EDR by cropping the spectra to three windows around H$\,\alpha$, H$\,\beta$ and H$\,\gamma$:
the wavelength ranges selected are $6500$--$6600\,\text{\AA}$; $4800$--$4900\,\text{\AA}$, and $4300$--$4400\,\text{\AA}$.
For each spectrum, a vector is created by cropping to each of these ranges, and concatenating the three `sub-vectors'.
Cropping the spectral vectors in this way gives $D=375$-dimensional vectors.
Following preprocessing and dimensionality reduction as above, the resulting embedding is shown in Fig.~\ref{fig:CVs}.
All 12 of the CVs identified by \citet{manser24} are located on a small island at the top right of the embedding, no doubt as a result of their shared emission features, most of which are double-peaked.
By zooming in to characteristic wavelength ranges, dimensionality reduction is therefore able to identify CVs from this sample with 100 per cent efficiency, according to human classification.

\subsection{Classifying new white dwarf spectra against DESI EDR}

\begin{figure*}
\includegraphics[width=\textwidth]{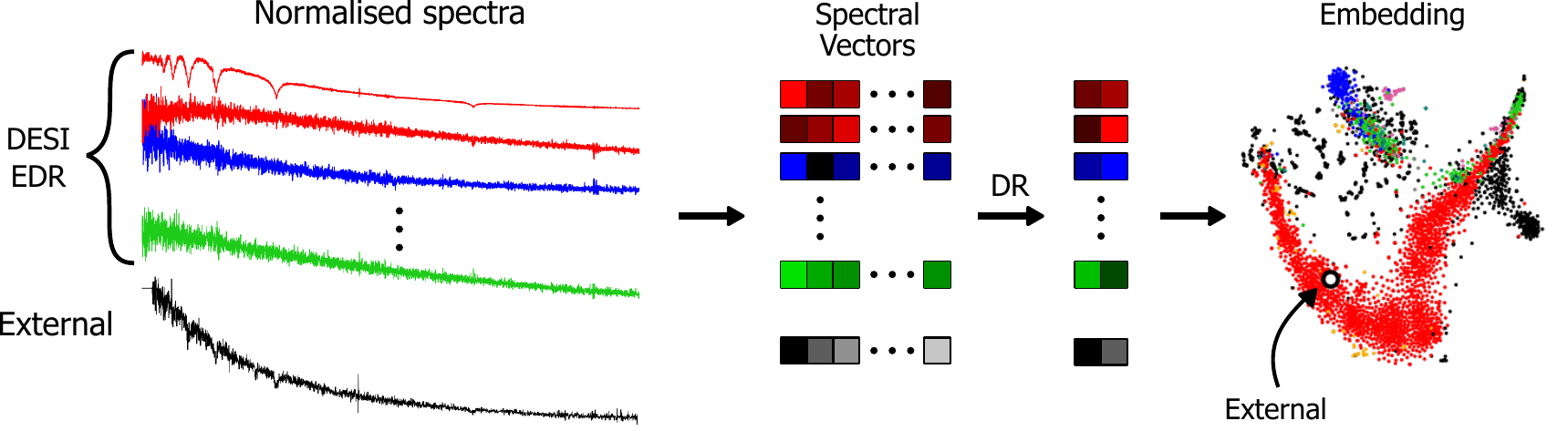}
\caption{
    Procedure for classifying external WD spectra against DESI EDR (see text).
    One would estimate the classification of this spectrum (black) as a DA.
}
\label{fig:external_procedure}
\end{figure*}

This work aims to provide a tool to help in the classification of imminent large spectroscopic WD datasets, rather than the classification of individual objects.
However, the latter can also be achieved, by using dimensionality reduction in a supervised way.
Here we present a means to estimate the classification of an additional spectrum against the $N$ WDs in the DESI EDR, by applying dimensionality reduction to a dataset of $N+1$ spectra: the DESI EDR WDs, plus the spectrum one wishes to classify.
This is achieved as follows, and as schematized in Fig.~\ref{fig:external_procedure}.
\begin{enumerate}
\item
Interpolate the external spectrum to the wavelength grid of the DESI spectra, and append it, giving a dataset of $N+1$ spectra: $\xb_1, \xb_2, \dots, \xb_N, \xb^* \in \mathbb{R}^{7781}$.
\item
Apply dimensionality reduction, giving $N+1$ two-dimensional points: $\yb_1, \dots,$ $\yb_N, \yb^* \in \mathbb{R}^2$.
Since $N\gg1$, the resulting embedding will not be noticeably different from the original embedding produced from the original set of $N$ spectra shown in Fig~\ref{fig:full_spectra}, though with an extra point $\yb^*$.
\item
Identify the location of the external spectrum $\yb^*$ in the new embedding.
The spectrum will be projected near to similar spectra, so its spectrum $\xb^*$ should be classified the same as the spectra projected near to it.
\end{enumerate}

\begin{figure*}
\includegraphics[width=\textwidth]{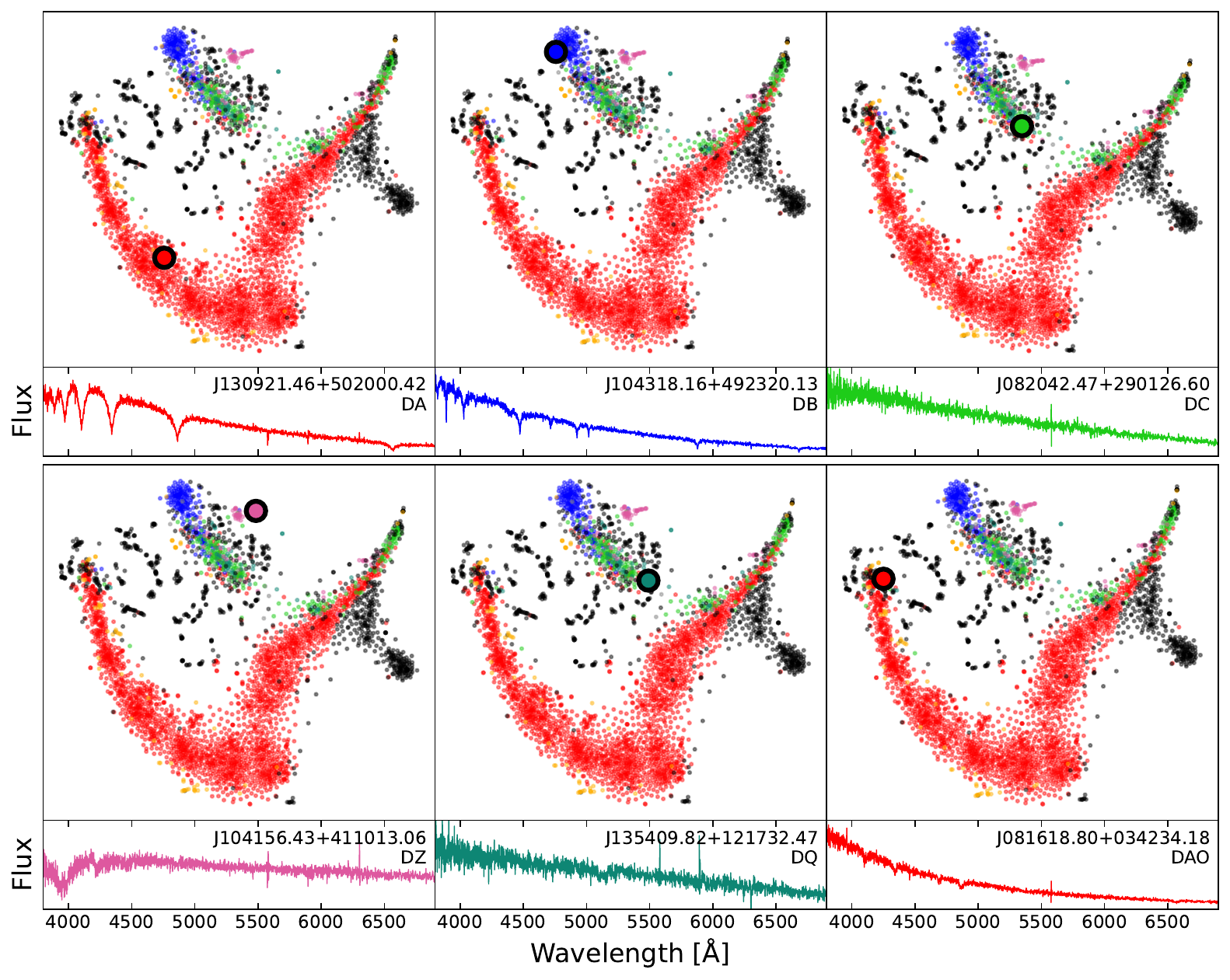}
\caption{
    Projection of external SDSS spectra appended to DESI EDR WD dataset.
    Each embedding is almost identical to the embedding of $N$ spectra shown in Fig.~\ref{fig:full_spectra}, but includes one additional point, corresponding to the external spectrum.
    The projection of the external spectrum is highlighted in each case, using the same colour scheme as Fig.~\ref{fig:full_spectra}.
    The DA, DB, and DZ are projected near to other objects classified as such, so these objects could reliably have been classified using dimensionality reduction as well as visual inspection.
    The DC, DQ, and DAO are also projected near to objects with the same class, but these regions of the embedding are more ambiguous.
    The external spectrum appended is shown in each case, together with its SDSS name and spectral classification (according to \citealt{gentilefusillo19}).
}
\label{fig:external_spectra}
\end{figure*}

To demonstrate this application, a spectrum from the SDSS WD catalogue compiled by \citet{gentilefusillo19} was selected at random from each of the following spectral classes: DA, DB, DZ, DC, DQ, DAO.
Each was interpolated to the same wavelength grid and then (individually) appended to the DESI EDR dataset described above.
$t$SNE was then applied to these six sets of $N+1$ spectra; the dimensionality reduction still only takes a few seconds.
The results of this procedure are shown in Fig.~\ref{fig:external_spectra}.

The SDSS DA is projected within the main DA sequence.
As such, if one did not \textit{a priori} know the classification of this spectrum, one could apply the above procedure, note that this spectrum is projected among other DAs, and be confident that this spectrum is also of a DA, having been identified by dimensionality reduction as a very similar spectrum.
Similarly, the SDSS DB and DZ are also projected into regions of the embedding with WDs with the correct classification.
The DCs, DQs and DAOs are also projected amongst other WDs in their respective classes, though these classes are not found in isolated islands even in the original embedding (Fig.~\ref{fig:full_spectra}).
For example, the DC is projected near to one end of the He-atmosphere sequence, which is occupied by not only DCs but also DQs, DZs, and even extragalactic sources.
Classifying them in the manner suggested would therefore be challenging.
Additional data would be necessary to distinguish it as a DC, including perhaps visual inspection.
However, this ambiguity is not necessarily a drawback.
As we outline in Section~\ref{sec:full_spectra}, the WD classification system does not separate WDs into wholly distinct classes: for example, a very cool DA is physically almost identical to a DC.
If an external WD spectrum is projected by the above method near the boundary between the cool DAs and the DCs, then this would inform the classifier that the spectrum is of a H-atmosphere WD around $5000\,\text{K}$.
Whether it should be classified as a DA or a DC has little physical meaning.

One could presumably apply this technique using the trick of cropping the spectra (Section~\ref{sec:zoom}), to investigate the similarity of a particular spectral range to those of the DESI EDR; this is beyond the scope of this work.

\section{Discussion}
\label{sec:discussion}

\subsection{An aid for spectral classification}

Approximately $70\,000$ WDs are targeted in the Milky Way Survey as part of the full DESI data release, expected in 2025 \citep{cooper23}.
Whereas the entire DESI EDR WD catalogue has been classified thanks to an extensive visual classification campaign \citep{manser24}, the full data release will contain a factor of 20 more WD candidate spectra.
We argue that automated methods such as dimensionality reduction would significantly reduce the expert time required to classify large spectroscopic surveys such as the DESI data release.

Under the assumption that the DESI EDR WD catalogue is representative of the WD population, applying dimensionality reduction to any large WD spectroscopic survey with a similar selection function would give maps analogous to those presented here.
Such maps would feature a large swathe of H-atmosphere DA*s and DCs organised by temperature, a second sequence of He-atmosphere DB*s, DQs, and DCs, and various islands of DZ*s, MS stars, and extragalactic sources.
Dimensionality reduction thus provides a powerful and rapid initial classification, broadly separating WDs with Balmer lines from those with He lines.
If a sufficiently diverse subset of the dataset is labelled -- as will be the case with the full DESI data release, given that the EDR has been classified -- then further classification can be done in a semi-supervised way.
Zooming in on particular lines, more precise classifications can be made with high recall.
Locating spectra on a map offers an intuitive -- and crucially, automated -- assessment of the likely classification of a spectrum.
This could be used in conjunction with other, non-spectral information, such as the source's \textit{Gaia} magnitudes, or the source's `probability of being a white dwarf' ($P_\text{WD}$; \citealt{gentilefusillo15}), when determining spectral class.

Dimensionality reduction offers several advantages over visual inspection, making it a useful complementary method.
Firstly, it is enormously quicker.
Secondly, as a result, it is far more readily repeatable and verifiable.
Thirdly, it is objective, as it deals with the data directly; visual inspection relies on the qualitative human interpretation of an image, which is less reproducible and permits no quantitative judgement of how much stronger a feature is than the noise.
Related to this point is the fact that dimensionality reduction effectively `sorts' the spectra by the strength of spectral features, as exhibited by the smooth transition from cool DAs to DCs (Section~\ref{sec:full_spectra}).

\subsection{Rare classes and the benefits of semi-supervised methods}
\label{sec:rare}

For some classes with little representation in the EDR, the full release may contain enough similar objects for them to form their own cluster.
For example, the EDR contains a single WD classified DO (J171600.53+422131.17; \citealt{manser24}), along with 10 DAOs, near the hot (left-hand) end of the DA sequence of Fig.~\ref{fig:full_spectra}.
With more DOs and DAOs in the full data release, it is reasonable to expect that dimensionality reduction would be able to separate them out from the DA sequence based on their common He~\textsc{ii} lines.

Such poorly-represented classes also highlight advantages of semi-supervised techniques over, for example, fully supervised machine learning.
The accuracy of supervised models is entirely dependent on their training data, and strongly imbalanced training sets make recognising underrepresented classes more difficult (e.g., \citealt{he09, johnson19, das23}).
As an extreme example, if one class constitutes 99 per cent of the training set, then a trivial model which predicts the dominant class for every data point would have a precision of 99 per cent.
Supervised machine learning techniques are thus less successful at identifying rarer classes.
With dimensionality reduction a substantially unique spectrum would be embedded far away from the more common classes.
Additionally, although multiclass classifiers allow for uncertainty in their predictions -- by outputting a set of numbers summing to one, often interpreted as a discrete probability distribution over the possible classes -- supervised classifiers often erroneously make very high-confidence predictions, especially for data not well-represented in the training set (e.g., \citealt{nguyen15, guo17, hein19}).
The semi-supervised technique presented in this work automatically accounts for such uncertain classifications, albeit qualitatively: spectra with minuscule Balmer features would be located in the overlap between DA and DC regions of the embedding, naturally informing a human classifier of some uncertainty in the classification.
As discussed in Section~\ref{sec:full_spectra}, this uncertainty is somewhat moot, and is as much a classification of the noise level in the spectrum as of the WD itself.
If a human classifier desired to distinguish between the two cases, they would devote more effort to identifying distinguishing features in the spectrum.
However, unambiguous spectra located solidly in the DA region would merely require a confirmatory glance at most.

\subsection{The role of effective temperature}
\label{sec:discteff}

Fig.~\ref{fig:full_spectra}(c) shows a trend in effective temperature from one end of the DA sequence to the other.
One might wonder whether preprocessing the spectra to remove temperature information would allow for more subtle trends to be exhibited in the embeddings.

It is difficult to remove all temperature information from a spectrum, such that all WD spectra of the same classification would appear the same.
Although the black-body continuum can be removed by fitting and subtraction, the shapes of WD absorption features are highly temperature-dependent (e.g., \citealt{liebert05, tremblay09}).
Additionally, the relative strengths of different spectral lines of the same species is temperature-dependent.
As such dimensionality reduction is still able to identify temperature trends in continuum-subtracted spectra.

\begin{figure*}
\includegraphics[width=\textwidth]{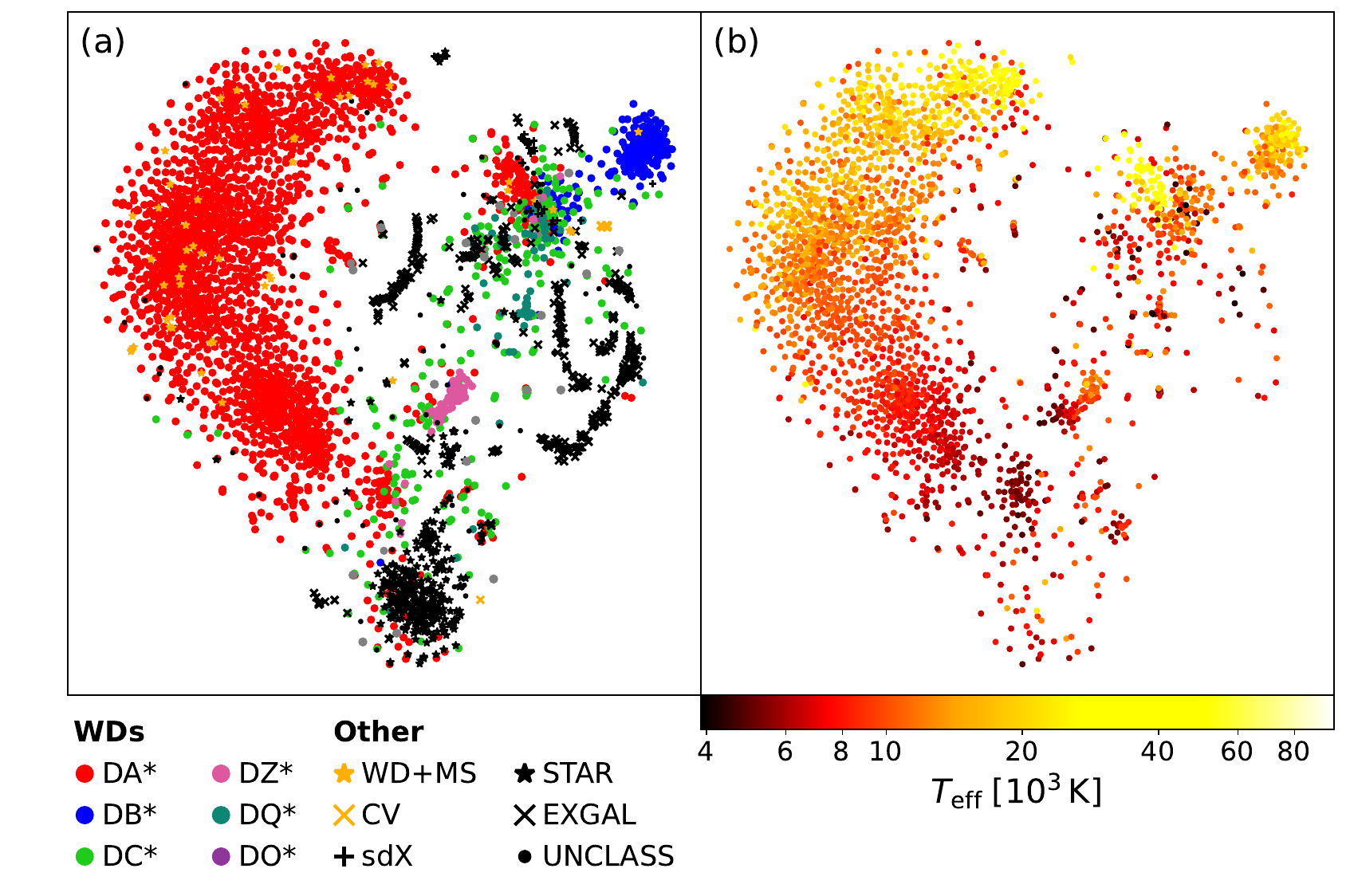}
\caption{
    Dimensionality-reduced embedding of spectra following continuum subtraction.
    (a) Colour-coded according to the visual classification of \citet{manser24}, as in Fig.~\ref{fig:full_spectra}(b).
    The DA sequence has partially collapsed, but not entirely.
    (b) Colour-coded by $\Teff$, as in Fig.~\ref{fig:full_spectra}(c).
    There is still a temperature trend along the shorter DA sequence.
}
\label{fig:continuum_subtracted}
\end{figure*}

This is illustrated in Fig.~\ref{fig:continuum_subtracted}, which shows the embedding of the spectra after continuum subtraction\footnote{
    Continuum subtraction was carried out using the \textsc{specutils} package, which by default fits a Chebyshev polynomial of degree 3.
}, preprocessing as above, and dimensionality reduction.
The DA sequence, which in Fig.~\ref{fig:full_spectra}(b) spanned the entire embedding, has collapsed somewhat.
This is indicative that the temperature information is no longer as prominent in the spectra after continuum subtraction, with other effects becoming important.
However, it is clear from Fig.~\ref{fig:continuum_subtracted}(b) that temperature information has not been completely removed, as there is still a trend from cool DA*s at the lower left to hot DA*s at the upper left.
This no doubt reflects the well-characterised differences in Balmer features between WDs of different temperatures.

We note in passing several other interesting features of this embedding that differ from the original embedding of non-continuum-subtracted spectra (Fig.~\ref{fig:full_spectra}).
\begin{itemize}
\item
DB*s are quite well-separated in the upper right of the embedding, more so than in the original embedding.
\item
Similarly, there is better separation of the MS stars, which form a cluster at the lower centre of the embedding.
\item
DZ*s are also much more well-clustered, with almost all being located in a cluster just below centre.
\item
All but one of the CVs is found in a small island just below the DB*s, even though the spectra have not been cropped (Section~\ref{sec:CVs}).
\end{itemize}

\subsection{Comparison with colour-colour diagrams}
\label{sec:colourcolour}

Colour cuts have often been used to select WDs from large photometric datasets such as SDSS \citep{eisenstein06, gentilefusillo15}, as well as to look for WDs with particular features such as metal pollution \citep{hollands15}.
These colour cuts are carried out by plotting each source in $(u-g)$-$(g-r)$ space and selecting some particular region.
In a sense, these colour-colour plots constitute a very rudimentary form of dimensionality reduction, as they deproject spectral information into a two-dimensional map.
Several similarities between these plots and the embeddings shown in this work are outlined here.

\begin{itemize}
\item
The embedding in Fig.~\ref{fig:full_spectra} is reminiscent of the $(u-g)$-$(g-r)$ colour-colour plot of WDs from the catalogue of \citet{gentilefusillo19}, shown in fig.~1 of \citet{manser20}.
The plot similarly shows a V-shaped sequence of DAs, with a secondary sequence of DBs, though they are considerably less well-separated.
This may not be entirely a coincidence.
As discussed above, the effect of $\Teff$ is to gradually `tilt' the spectrum, which would both continuously change the photometry, and also continuously transport the spectrum in the data space.
\item
Cool DBs, DCs, and DQs are all found in a similar location in the embedding of Fig.~\ref{fig:full_spectra}, as in $(u-g)$-$(g-r)$ space \citep{manser20}.
It is likely that the breadth and weakness of the spectral features of many instances of these classes has led them to be identified as similar spectra by dimensionality reduction.
Their proximity in colour space is due to their shared lack of strong Balmer features.
\item
In addition to being on a fairly isolated island in the embedding, cool DZs are also separated in colour space, due to strong Ca~\textsc{ii} absorption features in the SDSS \textit{u} band \citep{hollands15}.
\item
As discussed in Section~\ref{sec:full_spectra}, there is a small gap in the embeddings of DCs which fails to link the H- and He-atmosphere sequences.
There is a similar gap in the distribution of \textit{Gaia} colours of DCs in the DESI EDR around $\mathit{BP}-\mathit{RP}=0.75$ (see fig.~4 of \citealt{manser24}), suggesting poor sampling in a particular temperature region for DCs.
\end{itemize}

We argue that dimensionality reduction offers a significant improvement in the identification of spectral classes compared to simple colour-colour plots.
The separation between H- and He-dominated atmospheres is much more distinct, as the Balmer series seemingly has a larger effect on a spectrum's position in data space than on its photometry.
Further, while strong metal pollution can have a significant effect on photometry \citep{hollands15}, weaker absorption lines would not, yet dimensionality reduction readily identifies them.
This is evident in the fact that J160711.86+532157.6 is projected onto an island with other D*Zs, despite the Ca~\textsc{ii} lines being so weak that they were missed even by visual inspection (see Fig.~\ref{fig:DZisland}).
The increased power over colour-colour diagrams is ultimately due to the fact that while the presence of a narrow absorption line might have only a minimal impact on an object's photometry, it can move a spectral vector a significant distance in data space, allowing the line to be recognised by dimensionality reduction but not colour-colour plots.

\subsection{Limitations}

It is clear that dimensionality reduction alone is not capable of neatly classifying the spectra of all WD candidates.
Though the technique is quickly able to inform the classification of a large fraction of the dataset, particularly when focused on a wavelength range around some distinctive spectral features, drawbacks to the technique are outlined in this subsection.

As outlined in Section~\ref{sec:rare}, dimensionality reduction does not simply take a spectrum as input and produce a set of pseudo-probabilities of various classes, as with supervised multiclass classifiers.
While such a classification could be biased and may not reflect the nuances of WD spectral classification, the outputs are readily interpretable and require no further human intervention.
Locating a spectrum on a map relative to others may require some human judgement, which, while less time-consuming than examining the overall spectrum, is less trivial than simply receiving a set of probabilities and taking the class with the highest probability.

There are multiple regions of the embedding shown in Fig.~\ref{fig:full_spectra} where spectra of different classes are embedded close together.
The cooler end of the He-atmosphere sequence contains several different classes, mostly DCs, DQs, but also a few DB*s, DA*s, DZ*s, and extragalactic sources.
The spectra in this region show particularly shallow and broad features, such as the broad \ce{C2} Swan bands of the DQs.
This is likely due to the way dimensionality reduction handles noise.
As the different wavelength bins are each treated as vector components, there is no sense of `ordering': the technique is not aware of which wavelength bins are adjacent.
Dimensionality reduction does not compare different components of the same spectrum; rather it can only compare the values of a component between different spectra.
As such the method cannot account in for noise in the spectrum, and has difficulty in differentiating weak, broad spectral features because numerically the corresponding spectral vectors will not be very distant.
Conversely, very narrow spectral features, such as certain MS star absorption features, may also be difficult to distinguish, as the discrepancy may only be in a few wavelength bins.
If two spectra have very similar black-body continua, but one spectrum contains sharp features, the spectral vectors will likely be nearby.
This may explain the merging of the cool end of the H-atmosphere sequence with the MS stars towards the right of the embedding.
Dimensionality reduction is therefore best able to distinguish WD spectral classes where lines are neither very weak nor very narrow.

\subsection{Choices of spectral region}

Section~\ref{sec:DBs} showed that cropping the spectra to a region around a spectral line can improve dimensionality reduction's ability to determine whether WDs belong to particular classes.
It was found (by trial and improvement) that the ideal wavelength range should include both the spectral feature and a considerable amount of continuum, to give something for the spectral feature to `stand out against' following normalisation.
For example, the He feature at $5876\,\text{\AA}$ rarely has wings broader than $\pm50\,\text{\AA}$.
However, cropping the spectra to just this region led to the WDs with He lines being less distinctly separated from the rest of the spectra.

\subsection{Computation time}

Computation time has not been a significant issue in the dimensionality reduction of the DESI EDR WD catalogue ($5.5\,\text{s}$ for $N=3673$).
With the implementation of \textit{t}SNE used here, the complexity is $\mathcal{O}(N\log N)$ \citep{barneshut, vandermaaten14}, suggesting that for $N\approx 70\,000$ in the full data release the computation time would rise to $140\,\text{s}$.
The weak scaling makes dimensionality reduction a suitable tool for much larger spectroscopic datasets.

\section{Conclusions}
\label{sec:conclusions}

We outline the use of dimensionality reduction as an aid in the classification of large-sky spectroscopic surveys.
Providing a proof-of-concept through the application of $t$SNE to the DESI EDR WD catalogue, we demonstrate the method's ability to map out the structure of an intermediate-resolution spectroscopic survey, identifying spectra of various classes, in a way which naturally indicates uncertainty between classes.
By focusing on spectral windows containing particular features, sources with spectral features in these windows can be identified with high recall: CVs and WDs with helium lines can be identified with respectively 100 per cent and 90 per cent recall, as judged against human classification.
The technique identifies spectral features that have been missed even by visual classification, and takes only a few seconds to indicate spectral classifications for the entire catalogue.
Additionally, we present a means to use dimensionality reduction in a supervised manner to classify WD spectra by comparison with those in the DESI EDR.

The procedures outlined here could assist in the classification of the several upcoming large-sky spectroscopic surveys targeting WDs, enabling a quicker path to various astrophysical studies that depend on these classification.
These include: the evolution and internal structure of WDs; the behaviour and physical processes characterising CVs; the fraction of WDs showing pollution; the composition of exoplanetary material.
Such classification campaigns would proceed by applying dimensionality reduction to the spectroscopic dataset, labelling any whose spectral class is already known, and using this as an accurate prior for other spectra clustered nearby.
These methods would be equally applicable to the classification of other subsets of large spectroscopic surveys, such as main-sequence stars, quasars, or galaxies.
As the coverage, depth, and resolution of these surveys improve over the next few years, our work highlights the necessity and ease with which automated techniques can be exploited to maximise the scientific returns from these enormous datasets.

\section*{Acknowledgements}

The authors thank the anonymous reviewer for their many insightful comments, which have significantly improved the manuscript.
We also thank Andrew Swan for many useful comments, Keith Hawkins, Vasily Belokurov and Mariona Badenas-Agusti for discussions on the normalisation of spectra, Siyi Xu for advice on the interpretation of the embeddings, and Olivier Vincent for advice on nomenclature.

This research used data obtained with the Dark Energy Spectroscopic Instrument (DESI).
DESI construction and operations is managed by the Lawrence Berkeley National Laboratory.
This material is based upon work supported by the \href{https://www.energy.gov/}{U.S. Department of Energy}, Office of Science, Office of High-Energy Physics, under Contract No.\ DE--AC02--05CH11231, and by the National Energy Research Scientific Computing Center, a DOE Office of Science User Facility under the same contract.\
Additional support for DESI was provided by the \href{https://www.nsf.gov/}{U.S.\ National Science Foundation} (NSF), Division of Astronomical Sciences under Contract No.\ AST-0950945 to the NSF's National Optical-Infrared Astronomy Research Laboratory;
the \href{https://stfc.ukri.org/}{Science and Technology Facilities Council of the United Kingdom};
the \href{https://www.moore.org/}{Gordon and Betty Moore Foundation};
the \href{https://www.hsfoundation.org/}{Heising-Simons Foundation};
the \href{http://www.cea.fr/}{French Alternative Energies and Atomic Energy Commission} (CEA);
the \href{https://www.conacyt.gob.mx/}{National Council of Science and Technology of Mexico} (CONACYT);
the \href{http://www.mineco.gob.es/}{Ministry of Science and Innovation of Spain} (MICINN), and by the DESI Member Institutions: \url{www.desi.lbl.gov/collaborating-institutions}.
The DESI collaboration is honored to be permitted to conduct scientific research on Iolkam Du'ag (Kitt Peak), a mountain with particular significance to the \href{http://www.tonation-nsn.gov/}{Tohono O'odham Nation}.
Any opinions, findings, and conclusions or recommendations expressed in this material are those of the authors and do not necessarily reflect the views of the U.S. National Science Foundation, the U.S. Department of Energy, or any of the listed funding agencies.

Funding for the Sloan Digital Sky Survey IV has been provided by the Alfred P. Sloan Foundation, the U.S. Department of Energy Office of Science, and the Participating Institutions. 
SDSS-IV acknowledges support and resources from the Center for High Performance Computing at the University of Utah.
The SDSS website is \url{www.sdss4.org}.
SDSS-IV is managed by the Astrophysical Research Consortium for the Participating Institutions of the SDSS Collaboration including
the Brazilian Participation Group,
the Carnegie Institution for Science,
Carnegie Mellon University,
Center for Astrophysics | Harvard \& Smithsonian,
the Chilean Participation Group,
the French Participation Group,
Instituto de Astrof\'isica de Canarias,
The Johns Hopkins University,
Kavli Institute for the Physics and Mathematics of the Universe (IPMU) / University of Tokyo,
the Korean Participation Group,
Lawrence Berkeley National Laboratory,
Leibniz Institut f\"ur Astrophysik Potsdam (AIP),
Max-Planck-Institut f\"ur Astronomie (MPIA Heidelberg),
Max-Planck-Institut f\"ur Astrophysik (MPA Garching),
Max-Planck-Institut f\"ur Extraterrestrische Physik (MPE),
National Astronomical Observatories of China, New Mexico State University,
New York University,
University of Notre Dame,
Observat\'ario Nacional / MCTI,
The Ohio State University,
Pennsylvania State University,
Shanghai Astronomical Observatory,
United Kingdom Participation Group,
Universidad Nacional Aut\'onoma de M\'exico,
University of Arizona,
University of Colorado Boulder,
University of Oxford,
University of Portsmouth,
University of Utah,
University of Virginia,
University of Washington,
University of Wisconsin,
Vanderbilt University,
and Yale University.

This work has made use of data from the European Space Agency (ESA) mission \textit{Gaia} (\url{https://www.cosmos.esa.int/gaia}), processed by the \textit{Gaia} Data Processing and Analysis Consortium (DPAC,
\url{https://www.cosmos.esa.int/web/gaia/dpac/consortium}).
Funding for the DPAC has been provided by national institutions, in particular the institutions participating in the \textit{Gaia} Multilateral Agreement.

This research has made use of the SIMBAD database \citep{simbad} and the VizieR catalogue access tool \citep{vizier}, CDS, Strasbourg Astronomical Observatory, France.

In addition to Python packages referenced in the text, we also acknowledge the use of \textsc{NumPy} \citep{numpy}, \textsc{Matplotlib} \citep{matplotlib}, \textsc{PyVO} \citep{pyvo}, \textsc{pandas} \citep{pandas1, pandas2}, \textsc{SciPy} \citep{scipy}, \textsc{Astropy} \citep{astropy1, astropy2, astropy3}, and \textsc{bokeh} \citep{bokeh}.

\section*{Data Availability}

The data used in this work were obtained from the public archives of DESI, \textit{Gaia}, and SDSS.
Data products and Python scripts will be made available upon acceptance of the manuscript at \url{https://github.com/xbyrne/dr_wd_spectra}.
This repository will also contain interactive versions of the embeddings, where hovering over the points shows the corresponding spectra. 



\bibliographystyle{mnras}
\bibliography{bibliography} 




\appendix

\section{Mathematical details of \textit{t}SNE}
\label{sec:app}

The goal of dimensionality reduction is to map a set of $N$ high-dimensional vectors $\xb_1, \xb_2, \dots, \xb_N$ into two-dimensional vectors $\yb_1, \yb_2, \dots, \yb_N$, in such a way that the `similarity' between each pair of vectors is approximately preserved under the map.
Different dimensionality reduction methods use different definitions of similarity \citep{umap}; for $t$SNE, similarity between high-dimensional vectors $\xb_i$ and $\xb_j$ is defined by:
\begin{equation}
p_{ij}
= \frac{1}{2N} \qty(
    p_{i|j} + p_{j|i}
),
\end{equation}
where $p_{i|j}$ is defined by a normal distribution:
\begin{equation}
p_{i|j}
= \frac{
    \exp(-\norm{\xb_i - \xb_j}^2 / 2\sigma^2)
}{
    \sum_{k\neq i}^N
    \exp(-\norm{\xb_k - \xb_j}^2 / 2\sigma^2)
},
\end{equation}
where $\norm{\vdot}^2$ is the L2 norm and $\sigma$ is a hyperparameter known as the perplexity.
In this work, the perplexity is set to $\sigma=30$, the default value in the \textsc{scikit-learn} implementation used \citep{sklearn}.

Similarity between low-dimensional points $\yb_i$ and $\yb_j$ is defined by a student's $t$-distribution with one degree of freedom:
\begin{equation}
q_{ij}
= \frac{
    \qty(1 + \norm{\yb_i - \yb_j}^2)^{-1}
}{
    \sum_{k\neq i}^N
    \qty(1 + \norm{\yb_k - \yb_j}^2)^{-1}
}.
\end{equation}

To faithfully maintain the structure of the dataset as far as possible when reducing the dimensionality, the difference between the pairwise similarity distributions is minimised. The difference between the distributions is quantified by the Kullback-Leibler divergence \citep{kullbackleibler}:
\begin{equation}
\mathcal{KL}(p||q)
\equiv \sum_{i, j} p_{ij} \log \qty(
    \frac{p_{ij}}{q_{ij}}
),
\end{equation}
and is minimised by optimising the positions $\yb_i$ of the low-dimensional points, for example by gradient descent.
The result is a set of two-dimensional vectors $\yb_1, \yb_2, \dots \yb_N$ which are separated from each other by similar distances as $\xb_1, \xb_2, \dots \xb_N$ are from each other.

The \textsc{scikit-learn} implementation of $t$SNE used here \citep{sklearn} makes use of the Barnes-Hut algorithm \citep{barneshut}, speeding up the calculation of the embedding from $\mathcal{O}(N^2)$ to $\mathcal{O}(N\log N)$ at the expense of a very small reduction in accuracy \citep{vandermaaten14}.


\bsp	
\label{lastpage}
\end{document}